\documentclass[useAMS,usenatbib]{mn2e}

\usepackage{graphicx}
\usepackage{array}
\usepackage{amssymb}
\usepackage{color}
\newcolumntype{C}[1]{>{\centering\let\newline\\\arraybackslash\hspace{0pt}}m{#1}}

\title[Two bright $z$ \textgreater $\ $6 quasars from VST ATLAS]{Two bright $z$ \textgreater $\ $6 quasars from VST ATLAS and a new method of optical plus mid-infra-red colour selection}

\author[A. C. Carnall et al.]{A. C. Carnall$^{1}$\thanks{E-mail: a.c.carnall@durham.ac.uk}, T. Shanks$^{1}$\thanks{E-mail: tom.shanks@durham.ac.uk}, B. Chehade$^{1}$, M. Fumagalli$^{1,2}$, M. Rauch$^{2}$,\newauthor M. J. Irwin$^{3}$, E. Gonzalez-Solares$^{3}$, J. R. Findlay$^{1,4}$ and N. Metcalfe$^{1}$\\
$^{1}$Department of Physics, Durham University, South Road, Durham, DH1 3LE, UK\\
$^{2}$Observatories of the Carnegie Institution for Science, 813 Santa Barbara Street, Pasadena, CA 91101, USA\\
$^{3}$Institute of Astronomy, University of Cambridge, Madingley Rise, Cambridge, CB3 0HA, UK\\
$^{4}$Department of Physics \& Astronomy, University of Wyoming, 1000 E. University, Dept. 3905, Laramie, WY 82071, USA}

\begin{document}

\date{Accepted YYYY Month DD. Received YYYY Month DD; in original form YYYY Month DD}

\pagerange{\pageref{firstpage}--\pageref{lastpage}} \pubyear{2015}

\maketitle

\label{firstpage}

\begin{abstract}

\noindent We present the discovery of two $z$ \textgreater $\ $6 quasars, selected as i band dropouts in the VST ATLAS survey. Our first quasar has redshift, $z$ = 6.31 $\pm$ 0.03, z band magnitude, $z_{AB}$ = 19.63 $\pm$ 0.08 and rest frame 1450\r{A} absolute magnitude, $M_{1450}$ = -27.8 $\pm$ 0.2, making it the joint second most luminous quasar known at $z$ \textgreater $\ $6. The second quasar has $z$ = 6.02 $\pm$ 0.03, $z_{AB}$ = 19.54 $\pm$ 0.08 and $M_{1450}$ = -27.0 $\pm$ 0.1. We also recover a $z$ = 5.86 quasar discovered by Venemans et al. (2015, in prep.). To select our quasars we use a new 3D colour space, combining the ATLAS optical colours with mid-infra-red data from the Wide-field Infrared Survey Explorer (WISE). We use $i_{AB} - z_{AB}$ colour to exclude main sequence stars, galaxies and lower redshift quasars, $W1$ - $W2$ to exclude L dwarfs and $z_{AB} - W2$ to exclude T dwarfs. A restrictive set of colour cuts returns only our three high redshift quasars and no contaminants, albeit with a sample completeness of $\sim$50\%. We discuss how our 3D colour space can be used to reject the majority of contaminants from samples of bright 5.7 \textless $\ z$ \textless $\ $6.3 quasars, replacing follow-up near-infra-red photometry, whilst retaining high completeness. 

\end{abstract}

\begin{keywords}
quasars: general - quasars: individual: ATLAS J025.6821-33.4627 - quasars: individual: ATLAS J029.9915-36.5658
\end{keywords}

\section{Introduction}\label{sec1}

With the exception of transient events, bright quasars are the only objects luminous enough for high signal-to-noise ratio (SNR) spectra to be obtained out to very high redshifts. Such observations are key to our understanding of both the evolution of the intergalactic medium (IGM) during the epoch of reionization (e.g. \citealt{b37}) and the growth of the first supermassive black holes.

The Lyman alpha (Ly$\alpha$) forest observed in the spectra of quasars at moderate redshift ($z$ $\approx$ 3) arises from filaments of neutral hydrogen in the IGM. Models for reionization suggest that by $z$ $\sim$ 6, these filaments have started joining together to form a neutral IGM with high filling factor (e.g. \citealt{b2}, \citealt{b1}) resulting in the absorption of almost all flux at wavelengths blueward of Ly$\alpha$ in the quasar rest frame and causing the Gunn-Peterson troughs \citep{b33} observed in the spectra of quasars at these redshifts. 

Ionised regions at $z$ \textgreater $\ $6 are believed to be the direct result of photoionization by nearby AGN and star forming galaxies. Observations of these ``bubbles'' of ionised gas and the intervening neutral IGM, when illuminated by background quasars, can be used to probe the metallicity and ionised fraction of gas in the IGM, and in the vicinity of protogalaxies during the epoch of reionization. In the case of the current highest redshift known quasar, ULAS J1120+0641 at $z$ = 7.1 \citep{b3}, the gas in the vicinity of the object was shown to be significantly neutral and extremely metal poor (\citealt{b4}, \citealt{b5}), providing a first glimpse of the formation processes influencing protogalaxies at $z \sim 7$, and of the evolution and enrichment of the IGM during reionization. 

Furthermore, the apparent existence of $\sim 10^9\ \mathrm{M_{\odot}}$ black holes at this early point in cosmic history challenges our canonical pictures for the formation of black holes from accreting seeds, and the collapse of the dark matter haloes in which they reside. Two competing scenarios for massive black hole formation over such a short timescale exist (e.g. \citealt{b6}). Firstly, the direct collapse of vast quantities of warm gas may take place, a scenario which would require very specific conditions for collapse in the early Universe. Secondly, the merger of many smaller black holes, formed from the first generation of population III stars may take place within deep potential wells due to massive dark matter haloes ($\gtrsim10^{12}\ \mathrm{M_{\odot}}$), potentially leading to an anti-hierarchical distribution of black hole masses \citep{b7}.

Future progress in these fields will require high resolution spectroscopy of larger samples of $z$ \textgreater $\ $6 quasars. However, the spatial density of bright high redshift quasars is very low, with the luminosity function of \cite{b8} predicting only $\sim$110 $z$ \textgreater $\ $5.7 quasars with $z_{AB}$ \textless $\ $20.0 across the whole sky.

Conveniently, the sharp drop in flux blueward of Ly$\alpha$ due to IGM absorption allows high redshift quasar candidates to be selected from wide field optical survey data as dropouts at all wavelengths shorter than the z band for redshifts 5.7 \textless $\ z$ \textless $\ $6.5 and the Y band for objects with 6.5 \textless $\ z$ \textless $\ $7.2. However candidate lists obtained using this method, pioneered by \cite{b9}, are polluted by L and T dwarf stars \citep{b14} which are far more numerous than high redshift quasars.

This contamination means one or more further dimensions must be added to the colour space in order to differentiate between high redshift quasars and cool dwarfs. Initially follow-up J band photometry on a large number of i band dropout candidates was performed, e.g. by \citeauthor{b9} (\citeyear{b9} and later papers), \citeauthor{b12} (\citeyear{b12}, \citeyear{b13}) and \citeauthor{b8} (\citeyear{b10}, \citeyear{b11}, \citeyear{b8}). These authors select objects with blue colours in $z_{AB} - J$ to exclude L and T dwarfs.

Where near-infra-red photometry is available from survey data, a multidimensional colour space can be constructed without follow-up observations, e.g. \cite{b18} who use an $i_{AB} - z_{AB}$ vs $z_{AB} - Y$ colour space to select a population of 5.7 \textless $\ z$ \textless $\ $6.2 quasars, and \citeauthor{b15} (\citeyear{b15}, \citeyear{b3}) who use a statistical approach, described in \cite{b16}. Finally, \cite{b17} use a $z_{AB} - Y$ vs $Y - J$ colour space to select $z$ \textgreater $\ $6.5, z band dropout quasars. All of these methods however still require deep follow-up photometric observations in the i band to confidently exclude cool dwarf stars before proceeding to spectroscopic confirmation of their quasar candidates.

In this work, we propose a novel method of three dimensional optical plus mid-infra-red colour selection to define an extremely clean sample of 5.7 \textless $\ z$ \textless $\ $6.3 quasars, using data from the Wide-field Infrared Survey Explorer (WISE, \citealt{b19}), detailed in Section \ref{sec2.3}. We therefore show that follow-up near-infra-red observations can be replaced with publicly available mid-infra-red data to select the brightest i band dropout high redshift quasars in optical surveys. We then apply our method to the VST ATLAS survey, within which we select two new $z$ \textgreater $\ $6 quasars directly from the survey data without the need for imaging follow-up observations.

The structure of this paper is as follows. In Section \ref{sec2} we introduce the VST ATLAS survey, then set out the initial selection criteria we implemented to identify i band dropouts in our survey data. We then detail the optical plus mid-infra-red selection methods used to obtain a clean sample of $z_{AB}$ \textless $\ $20.0 high redshift quasars. In Section \ref{sec3} we report the discovery of two bright quasars selected from VST ATLAS survey data. In Section \ref{sec4} we discuss the utility of our selection method and how it can be more widely applied to select i band dropout quasars from optical survey data. We conclude and summarise our results in Section \ref{sec5}.

Throughout this paper, the following cosmological parameters are assumed: $H_{0} = 70\ \mathrm{km\ s^{-1}\ Mpc^{-1}}$, $\Omega_M$ = 0.28 and $\Omega_{\Lambda}$ = 0.72 \citep{b34}. All magnitudes are quoted on the AB system except for WISE magnitudes, $W1$ (3.4 $\mu m$) and $W2$ (4.6 $\mu m$) which are on the Vega system.

\begin{figure}
\includegraphics[width=0.48\textwidth]{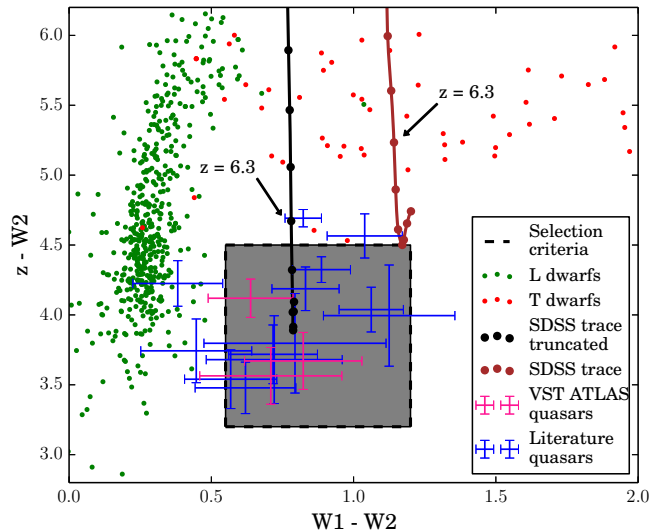}
\caption{Colour diagram for separating dwarf stars from high redshift quasars. The shaded region denotes our selection criteria. L and T dwarfs are plotted in green and red respectively. The quasar trace based solely upon the SDSS composite is plotted in brown and the truncated, power law extended trace in black (see Section \ref{sec2.3}). Both have points at $\Delta z = 0.1$ intervals, starting from $z$ = 5.7. The 13 published quasars with $z$ \textgreater $\ $5.7, $z_{AB}$ \textless $\ $20.0 and $W1$, $W2$ SNR \textgreater $\ 3.0$ are plotted in blue. The three high redshift quasars detected in VST ATLAS (Table \ref{table1}) are plotted in pink.}\label{fig1}
\end{figure}

\section{Candidate Selection}\label{sec2}

\begin{table*}
\begin{center}
\begin{tabular}{ C{1.4in} C{0.6in} C{0.7in} C{0.7in} C{0.9in} C{0.9in}}
\hline
 Quasar & Redshift & $i$ (AB mag) & $z$ (AB mag) & $W1$ (Vega mag) & $W2$ (Vega mag) \\
\hline
 ATLAS J025.6821-33.4627 & 6.31 $\pm$ 0.03 & \textgreater22.2 & 19.63 $\pm$ 0.06 & 16.12 $\pm$ 0.05 & 15.48 $\pm$ 0.09 \\
 ATLAS J029.9915-36.5658 & 6.02 $\pm$ 0.03 & \textgreater22.7 & 19.54 $\pm$ 0.08 & 16.70 $\pm$ 0.08 & 15.87 $\pm$ 0.13 \\
 VIKINGKiDS J0328-3253 & 5.86 $\pm$ 0.03 & \textgreater22.5 & 19.75 $\pm$ 0.12 & 16.90 $\pm$ 0.07 & 16.19 $\pm$ 0.13 \\
\hline
\end{tabular}
\end{center}
\caption{VST ATLAS and ALLWISE magnitudes for the three quasars selected from the VST ATLAS survey using the method described in Section \ref{sec2}. None of the objects are detected in the i band and 3$\sigma$ limiting magnitudes are provided.}\label{table1}
\end{table*}

\subsection{The VST ATLAS Survey}

The Very Large Telescope Survey Telescope (VST) ATLAS survey is an optical $ugriz$ survey on the 2.6m VST at Paranal, which aims to image $\sim$5000 square degrees of the Southern sky at similar depths to the Sloan Digital Sky Survey (SDSS) in the North. However VST ATLAS is deeper in the z band, with a mean 5$\mathrm{\sigma}$ limiting magnitude of 20.89 (as opposed to $\sim$20.5 for SDSS), and also has considerably better seeing, with median seeing in the range $\approx0.''8-1.''0$ for all five bands \citep{b35}.

All VST ATLAS data is processed by the Cambridge Astronomical Survey Unit (CASU) and the calibrated images and single band catalogues made available to the consortium and the Wide Field Astronomy Unit (WFAU) in Edinburgh. WFAU then produce band merged catalogues, available through the Omegacam Science Archive (OSA). ATLAS DR1, a public release containing data up to 30th September 2012 was delivered to ESO in November 2013.

\subsection{Initial Criteria}\label{sec2.2}

For this paper we use the ATLASv20131029 proprietary release, including data up to 31st March 2013 and containing 2060 square degrees imaged in the i and z bands. To produce an initial list of i band dropout objects within this area from the WFAU band merged catalogue, we specified that objects must have:

\medskip

\noindent\textbf{(i)} 18.0 \textless $\ z_{AB}$ \textless $\ $ 20.0. \\
\textbf{(ii)} Point source classification in the CASU catalogue (see Chehade et al. 2015, in prep. for a discussion of the robustness of this classification), and no WFAU error flags.\\
\textbf{(iii)} No catalogued detection in the i band (the catalogue detection limit is 5$\sigma$) or $i_{AB} - z_{AB}$ \textgreater $\ $2.2. \\
\textbf{(iv)} No catalogued detection in the u, g or r bands. \\

\noindent The upper limit imposed by (i) is necessary in order to confirm dropout status for objects undetected in the i band, the 5$\sigma$ i band detection limit for VST ATLAS being $i_{AB}$ = 22.0. The lower limit on magnitude imposed by (i) was based upon the prediction by the luminosity function of \cite{b8} that the spatial density of $z$ \textgreater $\ $5.7, $\ z_{AB}$ \textless $\ $ 18.0 quasars is less than one per 10,000 square degrees, therefore none were expected in our search area. Criterion (iv) excludes a very small number of sources which are assumed to be either due to imperfections in the catalogue band merging process or poor i band data in some fields.

After imposing these criteria on the data set, 9,607 objects remained in our sample. In order to exclude any spurious z band detections polluting the sample, and to facilitate our subsequent WISE colour selection process the following steps were then implemented. Objects must have:

\medskip

\noindent\textbf{(v)} A detection in the ALLWISE catalogue within 3."0. \\
\noindent\textbf{(vi)} SNR \textgreater $\ $3.0 in $W1$ and $W2$, corresponding to mean mag limits at -35$^{\circ}$  declination of $17.9$ in $W1$ and $17.0$ in $W2$. \\

\noindent After these criteria had been implemented, 3,452 objects remained in our sample. A visual inspection of a random sample of these candidates revealed significant contamination by objects which were either misclassified as i band dropouts due to imperfections in the band merging process or which could not be confirmed as i band dropouts due to areas of poor i band data.

To reject these objects, it is necessary to inspect cutout images of every candidate, however our sample is also still contaminated by L and T dwarfs. Our intention was to reject dwarfs using optical plus mid-infra-red colour selection and it was likely that a significant proportion of the other contaminants would also be rejected by this process. It was therefore decided to apply our selection to remove dwarf stars before inspecting cutout images for the full sample.

\begin{figure*}
\includegraphics[width=0.9\textwidth]{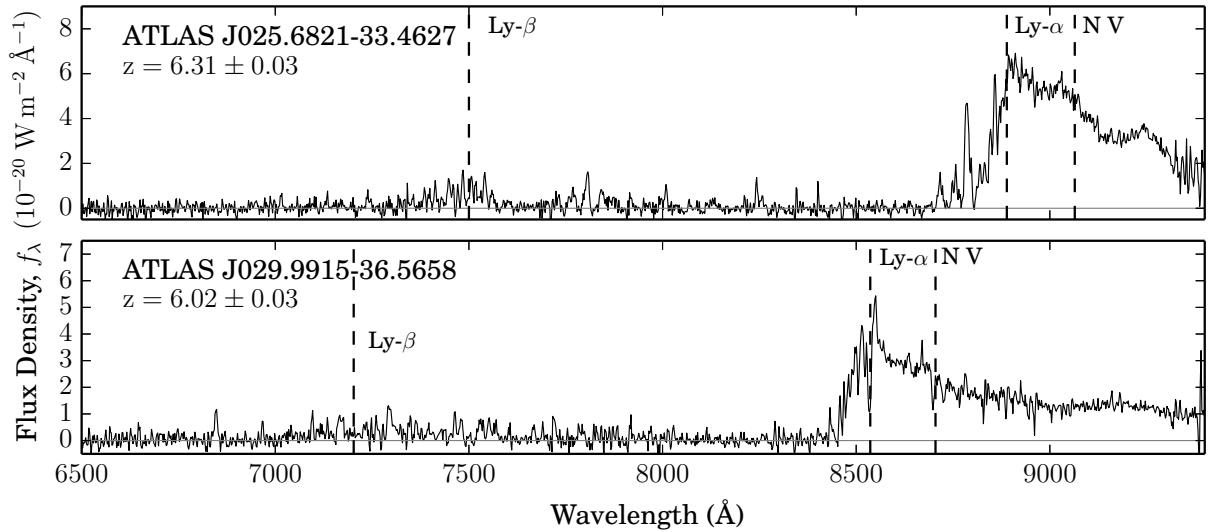}
\caption{LDSS-3 spectra for the two newly discovered quasars. The spectra have been binned to a resolution of 2\r{A}. Data reduction was carried out using the Astropy Python package \citep{b31}. Instrumental response was corrected using an observation of the standard star LTT 3864. Absolute flux calibration was carried out using synthetic photometry to scale to the z band magnitudes obtained from VST ATLAS. Positions consistent with our calculated redshifts are shown for the Ly-$\alpha$, Ly-$\beta$ and NV emission features.}\label{fig2}
\end{figure*}

\subsection{Optical Plus mid-infra-red Selection}\label{sec2.3}

It has been well documented that quasars have redder colour in $W1$ - $W2$ than main sequence stars (e.g. \citealt{b32}, \citealt{b20}). L dwarf stars also have bluer $W1$ - $W2$ colours, which we can exploit to reject L dwarfs from i band dropout samples by a $W1$ - $W2$ colour cut. T dwarfs inhabit a temperature range for which the black body components of their spectra peak in the mid-infra-red, and also exhibit temperature dependent methane absorption in the W1 band, giving them a broad range of redder $W1$ - $W2$ colours. However these objects are very dim by comparison in the optical, allowing us to split them off from high redshift quasars up to $z \sim 6.3$ by their very red $z_{AB} - W2$ colour.

In order to define suitable colour cuts, we first collected catalogues of L and T dwarfs in order to assess the distribution of their colours. We obtained samples of dwarf stars from \cite{b25}, \cite{b21}, \cite{b22}, \cite{b26}, \cite{b23}, \cite{b27} and \cite{b24} and matched them with the ALLWISE catalogue.

To understand the distribution of high redshift quasars in colour space we obtained magnitudes for all 17 published quasars with $z$ \textgreater $\ $5.7, $z_{AB}$ \textless $\ $20.0 (\citealt{b37}, \citealt{b15}, \citealt{b18} and \citealt{b38}). Of these, 13 have SNR greater than 3.0 in the W1 and W2 bands. We also produced a trace of the evolution of quasar colour with redshft by applying synthetic photometry to the median composite quasar spectrum of \cite{b28} (hereafter the SDSS composite).

IGM Ly$\alpha$ absorption was modelled by applying the corrections proposed by \cite{b29} between rest frame 1026\r{A} and 1216\r{A}. Flux blueward of Ly$\beta$ was assumed to be zero. The SDSS composite is known to be subject to greater host galaxy contamination at rest frame wavelengths above $\sim$5000\r{A} than is expected for highly luminous high redshift quasars. We attempted to mitigate this effect by truncating the SDSS composite at rest frame 4200\r{A} and extending to longer wavelengths using the bluer $\alpha_{\nu}$ = -0.44 power law used to fit the SDSS composite bluewards of 5000\r{A}. Figure \ref{fig1}  shows the $W1 - W2$ vs $z_{AB} - W2$ plane with the resulting trace plotted in black, along with the other data samples. A trace produced without truncation of the SDSS composite is shown in brown for comparison.

For this work, we defined a restrictive set of colour criteria based upon this information to select the best quasar candidates in our field. Our final criteria are as follows. Objects must have:

\medskip

\noindent\textbf{(vii)} 0.55 \textless $\ $ $W1$ - $W2$ \textless $\ $1.2 \\
\textbf{(viii)} 3.2 \textless $\ z_{AB}$ - $W2$ \textless $\ $4.5 \\

\noindent These colour criteria are also shown in Figure \ref{fig1}. It was found that $W1$ - $W2$ colour greater than 0.55 is a good criterion for excluding L dwarf stars, and $z_{AB}$ - $W2$ \textless $\ $ 4.5 discriminates well between T dwarfs and high redshift quasars up to $z \sim 6.3$. The upper limit on $W1$ - $W2$ and the lower limit on $z_{AB}$ - $W2$ were chosen so as to reject objects misclassified as i band dropouts and thus reduce the number of candidates to be visualy checked, without rejecting objects with colours similar to high redshift quasars. It must be stressed that the colour selection shown in Figure \ref{fig1} is only effective at selecting high redshift quasars if the sample being considered are all i band dropouts. The selection of Figure \ref{fig1} alone is not enough to exclude extragalactic contaminants such as lower redshift quasars.

After applying (vii) and (viii), 200 objects remained within our selection criteria. Upon visual inspection using the CASU postage stamp browser, 197 of these were found to be either obvious misclassifications for which the i band detection had not been successfully associated with the z band detection, or areas of poor i band data where dropout status could not be confirmed. Magnitudes for the three remaining sources are given in Table \ref{table1}. Of these sources, one was recognised as the only known $z$ \textgreater $\ $5.7, $z_{AB}$ \textless $\ $20.0 quasar in our search area, VIKINGKiDS J0328-3253 (Venemans et al. 2015, in prep.). As the luminosity function of \cite{b8} predicts six $z$ \textgreater $\ $5.7, $z_{AB}$ \textless $\ $20.0 quasars within our search area, we decided to proceed to spectroscopic confirmation of the two remaining candidates.

\section{Two New z \textgreater $\ $6 Quasars}\label{sec3}

\subsection{Spectral Observations}

We obtained long-slit spectra for our two candidates on the night of 16th January 2015 using the LDSS-3 instrument on the Magellan-II telescope at Las Campanas Observatory (observer M. Rauch). Exposures of 600s for each object were taken using the VPH RED grism and a $1."0$ slit. The data obtained are shown in Figure \ref{fig2}. Both objects show a strong spectral peak in the near-infra-red which we identify as Ly$\alpha$ emission, and a break in continuum flux blueward of this point. We now discuss the features of each spectrum in turn.

\subsection{ATLAS J025.6821-33.4627}

This quasar exhibits a broad Ly$\alpha$ emission line from which we have estimated a redshift, $z$ = 6.31 $\pm$ 0.03. By scaling the template spectrum employed in Section \ref{sec2.3} to the observed redshift and z band magnitude of the object we estimate it has $M_{1450}$ = -27.8 $\pm$ 0.2. This is the same as estimated for SDSS J1148+5251  by \cite{b37}, making this the joint second most luminous known $z$ \textgreater $\ $6 quasar, after SDSS J0100+2802 \citep{b38}.

The spectrum exhibits a faint signal consistent with Ly$\beta$, and an excess in transmission between 8700\r{A} and 8800\r{A}. This could indicate an ionised region close to the quasar, however near-infra-red spectroscopy is needed to better constrain the exact quasar redshift and hence the nature of any ionised regions.

\subsection{ATLAS J029.9915-36.5658}

For this quasar, we estimate a redshift of $z$ = 6.02 $\pm$ 0.03 based upon the position of the broad Ly$\alpha$ line. Following the same procedure as above, we estimate an absolute magnitude of $M_{1450}$ = -27.0 $\pm$ 0.1. There is a probable absorption feature superimposed on the Ly$\alpha$ emission line, as can also be observed in the spectra of several of the quasars first reported by \cite{b18}.

\section{Discussion}\label{sec4}

In this work we have employed a restrictive set of colour criteria. This approach was chosen in order to prove the concept of our optical plus mid-infra-red selection method by reducing contamination to a very low level. Naturally this takes a toll on the completeness of the sample of high redshift quasars obtained. The luminosity function of \cite{b8} predicts six quasars with redshift $z$ \textgreater $\ $5.7 and z band magnitude $z_{AB}$ \textless $\ $20.0 over the 2060 square degrees of our search area. The confirmation of all three of our candidates as high redshift quasars confirms the extremely clean nature of our colour selection, and places the completeness of our sample at $\sim$50\%. This is supported by the fact that 9 of the 17 published $z$ \textgreater $\ $5.7, $z_{AB}$ \textless $\ $20.0 quasars fall into our selection criteria, suggesting a completeness of $\sim$53\%.

When considering how to extend our colour selection to obtain a more complete sample of quasars, it is instructive to analyse the location in colour space of the eight quasars which do not meet our selection criteria as plotted in Figure \ref{fig1}. Four quasars are not detected in the ALLWISE source catalogue within 3."0 with SNR above 3.0 in the $W1$ and $W2$ bands. A further two fall to the left of our selection box, with $W1$ - $W2$ colours of 0.38 and 0.44, closer to the distribution of L dwarf stars. Two other objects fall just above our selection box with $z_{AB} - W2$ = 4.56 and 4.69.

Our selection criteria could be extended and coupled with follow-up observations in the optical to select a more complete population of bright 5.7 \textless $\ z$ \textless $\ $6.3 quasars. A natural method for this would involve relaxing the lower bound on $W1$ - $W2$ and upper bound on $z_{AB} - W2$, then conducting follow-up i band observations to exclude L and T dwarfs within the extended selection region. This selection still results in the removal of the majority of L and T dwarfs, meaning a significant reduction in the number of candidate objects for which follow-up observations are required.

Our method has two limitations, the first of which is the requirement that candidates be bright enough in the mid-infra-red to appear in the $W1$ and $W2$ bands with a SNR greater than 3.0. Currently, this limits the usefulness of our method in selecting lower luminosity quasars, however ongoing and future deeper surveys in the mid-infra-red such as the Spitzer-IRAC Equatorial Survey (SpIES) and NEOWISE will soon allow this method to be applied to objects with lower intrinsic luminosities. Secondly the steep increase of the $z_{AB} - W2$ colour of quasars with redshift above $z$ $\sim$ 6.2 means we cannot select clean samples of quasars with $z$ \textgreater $\ $6.3 using our method. An analogous colour selection for higher redshift objects using $Y - W2$ colour fails to efficiently separate T dwarfs from high redshift quasars.

\section{Conclusion} \label{sec5}

In summary we have proven the concept of a new method of optical plus mid-infra-red colour selection for 5.7 \textless $\ z$ \textless $\ $6.3 quasars. Using a restrictive set of colour criteria we identified two candidates which were confirmed to be bright quasars at redshifts of 6.31 $\pm$ 0.03 and 6.02 $\pm$ 0.03, with $M_{1450}$ = -27.8 $\pm$ 0.2 and -27.0 $\pm$ 0.1 respectively. The former is the joint second most luminous known $z$ \textgreater $\ $6 quasar and both are well positioned for observation by powerful current and future Southern observatories. Our method allows selection of i band dropout quasars to proceed from optical catalogues with a significant reduction to the volume of follow-up observations by replacing Y or J band follow-up photometry with publicly available mid-infra-red WISE data.

\section*{Acknowledgments}

This research made use of Astropy, a community-developed core Python package for Astronomy \citep{b31}. This paper is based on observations obtained as part of the VST ATLAS Survey, ESO Program, 177.A-3011 (PI: Shanks). The UK STFC is acknowledged for postdoctoral support for J. R. Findlay, PhD studentship support for B. Chehade and for support for the Cambridge Astronomical Surveys Unit and the Wide-Field Astronomy Unit. This publication makes use of data products from the Wide-field Infrared Survey Explorer, which is a joint project of the University of California, Los Angeles, and the Jet Propulsion Laboratory/California Institute of Technology, funded by the National Aeronautics and Space Administration. This paper also includes data gathered with the 6.5 metre Magellan Telescopes located at Las Campanas Observatory, Chile.

\appendix

\label{lastpage}

\end{document}